\documentclass[12pt]{article}
\usepackage{amssymb,amsmath,amsthm,graphicx,ulem}

\pdfoutput=1

\usepackage{graphicx,subfigure}
\usepackage{epsfig}
\usepackage{amsmath}
\usepackage{amsfonts}
\usepackage{amssymb}
\usepackage[usenames]{color}
\usepackage[letterpaper,left=2.cm,right=2.cm,top=2.5cm,bottom=2.5cm]{geometry}
\usepackage{float}

\newcommand{\beq}{\begin{equation}}
\newcommand{\eeq}{\end{equation}}
\newcommand{\be}{\begin{equation}}
\newcommand{\ee}{\end{equation}}
\newcommand{\beqa}{\begin{eqnarray}}
\newcommand{\eeqa}{\end{eqnarray}}
\newcommand{\beqar}{\begin{eqnarray*}}
\newcommand{\eeqar}{\end{eqnarray*}}
\newcommand{\bea}{\begin{eqnarray}}
\newcommand{\eea}{\end{eqnarray}}

\numberwithin{equation}{section}
\begin{document}
 
\allowdisplaybreaks
\normalem
\title{Geons and Spin-2 Condensates in the AdS Soliton}
\author{Gavin S. Hartnett and Gary T. Horowitz
\\ \\ 
Department of Physics, UCSB, Santa Barbara, CA 93106, USA 
\\ 
\small{hartnett@physics.ucsb.edu, gary@physics.ucsb.edu}}
\date{}
\maketitle 

\begin{abstract}
We construct geons starting with gravitational perturbations of the AdS soliton. Previous studies of a charged scalar field in the soliton background showed a holographic insulator/superconductor transition at a critical chemical potential. We explore the possibility that dimensional reduction of  the geon could model a transition to a $d$-wave superconductor. We find that although one does get a charged spin-2 condensate, it has higher free energy than the state without the condensate, so there is no phase transition.
\end{abstract}

\newpage

\baselineskip16pt

\section{Introduction}
A geon is a classical solution to the vacuum Einstein equations representing a localized and non-dispersing lump of gravitational energy held together by its own gravitational attraction.  Geons were originally  conjectured to exist by Wheeler, and the first serious attempt to construct them was made by Brill and Hartle \cite{BrillHartle}.  Their geon, however, has only a finite lifetime; at late times the gravitational waves comprising the geon will break free and disperse. This is a common feature of asymptotically flat geons \cite{Louko}. The tendency of geons to disperse can be remedied with a negative cosmological constant, as anti-de Sitter (AdS) space acts like a confining box. Perturbative geons in global AdS in four dimensions have been recently constructed in \cite{Dias1}. This provides strong evidence that associated with every individual linearized graviton mode in global AdS, there is a one parameter family of exact nonsingular geons\footnote{This is not true for generic perturbations of AdS containing superpositions of modes \cite{Dias1,Bizon}.}.

One should be able to construct geons starting with gravitational perturbations of any locally asymptotically AdS ground state. In particular, one can start with the AdS soliton \cite{Witten,HorowitzMyers}, which has seen several applications in gauge/gravity duality. It was originally introduced to describe the ground state of a confining gauge theory \cite{Witten}, but in more recent condensed matter applications it has been used to construct the gravitational dual of an  insulator/superconductor quantum phase transition \cite{Nishioka,HorowitzWay}.
 In this note we perturbatively construct a class of geons starting with the five-dimensional AdS soliton. 

Although the geons we find are interesting in their own right, perhaps the most intriguing aspect of these solutions is the potential connection with a $d$-wave superconductor. The possibility of obtaining a gravitational dual of a $d$-wave superconductor is exciting because this is what is seen in the high-$T_c$ cuprates. While the original holographic superconductor exhibited an $s$-wave order parameter \cite{HHH}, and later a model with a $p$-wave order parameter was constructed in \cite{GubserPufu}, the $d$-wave case  remains elusive. The major obstacle to building a $d$-wave superconductor is that there is no known consistent action for a charged, massive, spin-2 field minimally coupled to gravity. Various authors \cite{Chen,Benini} have worked with incomplete actions and found $d$-wave superconducting condensates, but as of yet no consistent holographic model has been found.

An alternative approach towards constructing a $d$-wave holographic superconductor is to study metric perturbations in Kaluza-Klein theory. Upon dimensional reduction, gravitons carrying momentum along a compact direction become charged under a gauge group that corresponds to the isometry group of the internal manifold.  Kaluza-Klein gravity therefore provides a natural framework to find a consistent theory of charged spin-2 fields coupled to gravity. We will explore the feasibility of this approach using the  geons we construct\footnote{During this work we learned that the Kaluza-Klein approach to $d$-wave superconductivity is also being studied by Kim et al \cite{Kim}, who consider dimensionally reduced supergravity in 10 and 11 dimensions.}. 

The AdS soliton is a particularly interesting background to consider because in addition to the $U(1)$  gauge field coming from the dimensional reduction along its $S^1$, the linearized metric perturbations may be decomposed into scalar, vector, and tensor types, corresponding to $s$, $p$, and $d$-wave excitations in the boundary theory. The Kaluza-Klein approach applied to the AdS soliton thus provides the exciting possibility of being able to describe a range of qualitatively different superconductors using just pure gravity. Unfortunately, we find that $d$-wave superconductors are not described by the perturbative geons we construct because the condensates are never thermodynamically preferred. This result is in contrast to the one obtained when one puts a Maxwell field and charged scalar field in the soliton background (rather than obtain them by Kaluza-Klein reduction). In that case, there is a continuous phase transition which turns on the scalar field as one increases the chemical potential. We discuss a likely explanation for the different behavior.

This paper is organized as follows. In section two the linearized metric perturbations of the AdS soliton are reviewed and the perturbative construction of the geons is described. In section three holographic superconductors based on the AdS soliton background are briefly reviewed, and the possibility that the dimensionally reduced geons could describe $d$-wave superconductors is considered. 

\section{Geons Built from the AdS Soliton }
\subsection{Linearized Metric Perturbations}
We study five-dimensional gravity described by the simple action
\be 
S = \int d^5x \sqrt{-g} \Big(R+\frac{12}{L^2}\Big),
\ee
where 
we have set $16\pi G = 1$. The AdS soliton is a solution of the above action with line element
\be 
ds^2 = \frac{L^2}{r^2 f(r)}dr^2 + \frac{r^2}{L^2}\Big(f(r)d\tilde y^2 + \sum_{\mu,\nu=0}^2 \eta_{\mu\nu} dx^{\mu}dx^{\nu} \Big), \qquad f(r) = 1 - \Big(\frac{r_0}{r}\Big)^{4}. 
\ee
This solution can be obtained from the planar black hole via a double Wick rotation. The geometry smoothly caps off at $r=r_0$ if the $\tilde y$-coordinate is chosen to be periodic with period $ \pi L^2/r_0$. 
The conformal boundary is the direct product of three-dimensional Minkowski spacetime with a circle, $\mathbb{M}^{3}\times S^1$.

Throughout this paper we work in perturbation theory around the background of the AdS soliton. For this purpose an expansion parameter $\epsilon$ is introduced and the metric is expanded as
\be 
g_{AB}(\epsilon) = \sum_{i=0}^{\infty} \epsilon^i g_{AB}^{(i)}, 
\ee 
where $g_{AB}^{(0)}$ is the AdS soliton metric.  Since the Einstein equations will need to be solved numerically, it will be useful to work with a compactified radial coordinate and to scale the dimensionful constants out of the metric. The background line element then becomes
\be 
g^{(0)}_{AB}dx^A dx^B = \frac{1}{z^2}\Bigg[\frac{dz^2}{(1-z^4)} + \frac{1}{4}(1-z^4)dy^2 -dt^2 + dx_1^2 + dx_2^2 \Bigg]. 
\ee
The circle coordinate has been rescaled so that $y\sim y + 2\pi$. The radial coordinate takes values in the range $z \in [0,1]$. The conformal boundary is at $z=0$ and the tip is located at $z=1$.

Next, we briefly review linearized metric perturbations of the AdS soliton background. A comprehensive treatment may be found in \cite{ConstableMyers}. The perturbations are decomposed into scalar, vector, and tensor types according to their transformation properties under the $SO(1,2)$ Lorentz symmetry of the soliton. We shall focus on  tensor perturbations, which can be parametrized as
\be\label{graviton} 
g^{(1)}_{AB} = \varepsilon_{AB} \frac{H(z)}{z^2} \cos(qy - \omega t).
\ee
Here $q \in \mathbb{Z}$ and $\varepsilon_{AB}$ is a polarization tensor. In five-dimensions, there are two independent polarizations,
\be\label{pol}
\varepsilon_{x_1 x_2} = 1, \qquad \text{all other components zero},
\ee
\be 
\varepsilon_{x_1 x_1} = - \varepsilon_{x_2 x_2} = 1, \qquad \text{all other components zero}.
\ee

With the above ansatz for the metric perturbation, the Einstein equations reduce to a single second order ODE for the function $H(z)$. Requiring that the perturbation be regular at the tip and normalizable at the boundary yields a Sturm-Liouville eigenvalue problem. Solutions only exist for a discrete set of frequencies $\omega \in \mathbb{R}$. These solutions are the normal modes and the $\omega$ are the normal mode frequencies. The tensor perturbations may therefore be characterized by two integers, the radial overtone number and the momentum $q$.

The perturbation breaks some, but not all, of the symmetries of the background. Translational invariance in the $t$ and $y$ directions is broken, but the perturbation remains invariant under the helical Killing field $K=\partial_t + (\omega/q) \partial_y$. Since $\omega > 2q$, this Killing field is timelike near the tip and spacelike near the boundary. The perturbation also remains invariant under translations in the $x_1, x_2$ directions and under the combined discrete operations of $t\rightarrow -t$ and $y\rightarrow-y$.
\subsection{Perturbative Construction of   Geons}
We now extend the study of the previous section to higher orders in perturbation theory. We will call the resulting metrics geons as they are solutions describing nonlinear, non-dispersing concentrations of gravitational waves. The structure of the perturbative construction is very similar to Gubser's study of the non-uniform black string \cite{Gubser}. For every  distinct linear perturbation there should exist  a corresponding one-parameter family of geons, $g_{AB}(\epsilon)$, with symmetries similar to those of the linearized mode. The expansion parameter will be chosen so that the momentum of the geon is $P = \epsilon^2 V_2$, where $V_2$ is the coordinate volume of the $x_1$-$x_2$ plane, although other choices are also possible. The metric ansatz is
\be 
ds^2 = \frac{1}{z^2}\Bigg[Ady^2 + Bdz^2 + Cdt^2 + D(dx_1^2+dx_2^2) + E dt dy + F dz dt + G dz dy  + H dx_1 dx_2 \Bigg]. 
\ee

This ansatz corresponds to a geon seeded by a tensor perturbation with polarization (\ref{pol}). Here $A$ through $H$ are functions of the coordinates $z,t$, and $y$, as well as the expansion parameter $\epsilon$. Periodicity of the circle coordinate implies that the $y$-dependence of these functions will be organized into a Fourier series, while the existence of the helical Killing vector implies that the functions will depend on $y$ and $t$ only through the combination $(q y -\omega t)$. Thus the expansions for $A$ through $E$ take the form
\be 
A(z,y,t,\epsilon) = \sum_{n=0}^{\infty} A_{2n}(z,\epsilon) \cos\Big(2n(q y - \omega t)\Big),
\ee
while the expansions for $F$ and $G$ take the form
\be 
F(z,y,t,\epsilon) = \sum_{n=1}^{\infty} F_{2n}(z,\epsilon) \sin\Big(2n(qy -\omega t)\Big).
\ee
Whether a given metric function is written as a sum of sines or cosines is determined by requiring that the line element be invariant under the discrete symmetry $(y,t) \rightarrow (-y,-t)$. The amplitude of each Fourier mode may also be expanded in powers of $\epsilon$. The expansions for $A_{2n}$ through $G_{2n}$ take the form
\be 
A_{2n}(z,\epsilon) = \sum_{i=0}^{\infty} \epsilon^{2n+2i}A_{2n}^{(2i)}(z).
\ee
The structure of the perturbative equations dictates that only even Fourier modes and even powers of $\epsilon$ appear in the above expansions. The function $H$ differs from functions $A$ through $G$ because it seeds the geon.  It is expanded in odd Fourier modes and in odd powers of $\epsilon$,
\be 
H(z,y,t,\epsilon) = \sum_{n=0}^{\infty}H_{2n+1}(z,\epsilon)\cos\Big((2n+1)(qy-\omega t)\Big),
 \qquad H_{2n+1}(z,\epsilon) = \sum_{i=0}^{\infty}\epsilon^{2n+2i+1} H_{2n+1}^{(2i)}(z).
 \ee
Note that although an infinite number of Fourier modes will be needed to describe the solution, the higher modes only enter  at higher orders in perturbation theory. 

 When constructing geons in global AdS, there was a potential problem due to resonances. Since all the normal mode frequencies were integer multiples of the lowest one, there were resonances at higher orders in the perturbative construction which threatened to introduce growing modes into the solution. It was found however, that  this could be avoided by letting $\omega$ depend on $\epsilon$, provided one starts with individual graviton modes \cite{Dias1}. 
 
 The normal mode frequencies in  the AdS soliton are not integer multiples of each other, so there are no exact resonances. Nevertheless $\omega$ still depends on $\epsilon$ for the following reason: recall that for the linearized perturbation, the value of $\omega$ was determined by requiring that $H(z)$ be regular at the tip and normalizable at the boundary. This linear perturbation then seeds the nonlinear geon, and is relabelled $H_1^{(0)}(z)$. At second order, the background geometry is altered, and at third order, the original seed Fourier mode is corrected because the function $H_{1}^{(2)}(z)$ is turned on. The original value for $\omega$ does not allow $H_{1}^{(2)}(z)$ to be both regular and normalizable, therefore $\omega$ must also be corrected. These corrections will be determined from odd orders in perturbation theory, but the corrections to $\omega$ will be in even powers of $\epsilon$,
\be 
\omega(\epsilon) = \sum_{i=0}^{\infty}\epsilon^{2i}\omega_{2i}.
\ee

The geon is not invariant under the action of the vectors $\partial_t$ or $\partial_y$ individually, but these are asymptotic Killing vectors and therefore the geon will carry the corresponding conserved charges of energy and momentum. These charges also have an expansion in even powers of $\epsilon$,
\be 
E(\epsilon) = \sum_{i=0}^{\infty}\epsilon^{2i} E_{2i}, \qquad P(\epsilon) = \sum_{i=0}^{\infty}\epsilon^{2i} P_{2i}. 
\ee
In the above expansions, $\epsilon$ appears as a formal expansion parameter. In order to relate the expansion parameter to a physical quantity, we define $\epsilon$ to make $P(\epsilon) = \epsilon^2 V_2$. 

At each order in perturbation theory the Einstein equations reduce to a set of coupled ODE's. The first order equations are homogeneous, while all the higher order equations have inhomogeneous source terms consisting of powers of the lower order functions and their derivatives. We choose boundary conditions that enforce regularity at the tip and leave the boundary metric unchanged. We also fix the periodicity of the circle to be $2\pi$. 

We numerically solved the perturbative hierarchy of Einstein's equations to third order. The energy and momentum were obtained using the AdS stress tensor formalism of Ref.  \cite{Balasubramanian}. We find that the geon has larger energy than the background, $E_2 > 0$, in accord with the  positive energy conjecture \cite{HorowitzMyers}. As a numerical check, we verified that the solutions obey the first law of thermodynamics to $\mathcal{O}(\epsilon^4)$. Since there are no horizons (and hence no entropy) and the solution is invariant under $\partial_t + \omega/q\ \partial_y$, this is simply \cite{Wald:1993ki}
\be\label{firstlaw} 
dE = \frac{\omega}{q}dP.
\ee
 We also checked that the boundary stress tensor is traceless. Odd-dimensional asymptotically locally AdS spacetimes may have a conformal anomaly, but not when the boundary is Ricci flat, as it is for these geons. Since there are no exact resonances, there should be no obstructions to extending the perturbative solution to all orders and constructing a one parameter family of exact solutions for each linearized mode.

For later reference, we now consider how the frequency changes with the Kaluza-Klein excitation $q$: $\omega(q) = \omega_0(q) + \omega_2(q)\epsilon^2 + \mathcal{O}(\epsilon^4)$. For large $q$, one finds $\omega_0 \rightarrow 2 q$. This is a simple consequence of the fact that as $q\rightarrow \infty$, the support of the first order normal mode becomes both squeezed (in the coordinate $z$) and pushed further away from the tip (see Fig. 1). For arbitrarily large $q$, the normal mode is effectively localized on the boundary, and so its equation of motion becomes that of a massless scalar field on $\mathbb{M}^3\times S^1$. This leads to the dispersion relation $\omega_0 = 2 q$. The factor of 2 arises from the fact that the $y$ coordinate has been rescaled to make its period $2\pi$. As discussed above, the second order term in the frequency expansion, $\omega_2$, is found by requiring that the function $H_1^{(2)}(z)$ be regular at the tip and normalizable at the boundary. The result is always negative (see Fig. 2). With our choice of expansion parameter,  $P=\epsilon^2 V_2 $, the second order change asymptotes to $0$ as $q\rightarrow \infty$. This is simply a reflection of the fact that as $q$ gets larger, the amplitude of the linearized perturbation must decrease to maintain a fixed $P$. Thus the second order backreaction becomes smaller and hence there is a smaller change in the frequency. The fact that $\omega_2 < 0$ will play an important role in the next section.   The global AdS geons also had $\omega_2 < 0$ \cite{Dias1}. 

\begin{figure}[H]
\centering
\includegraphics[scale=.8]{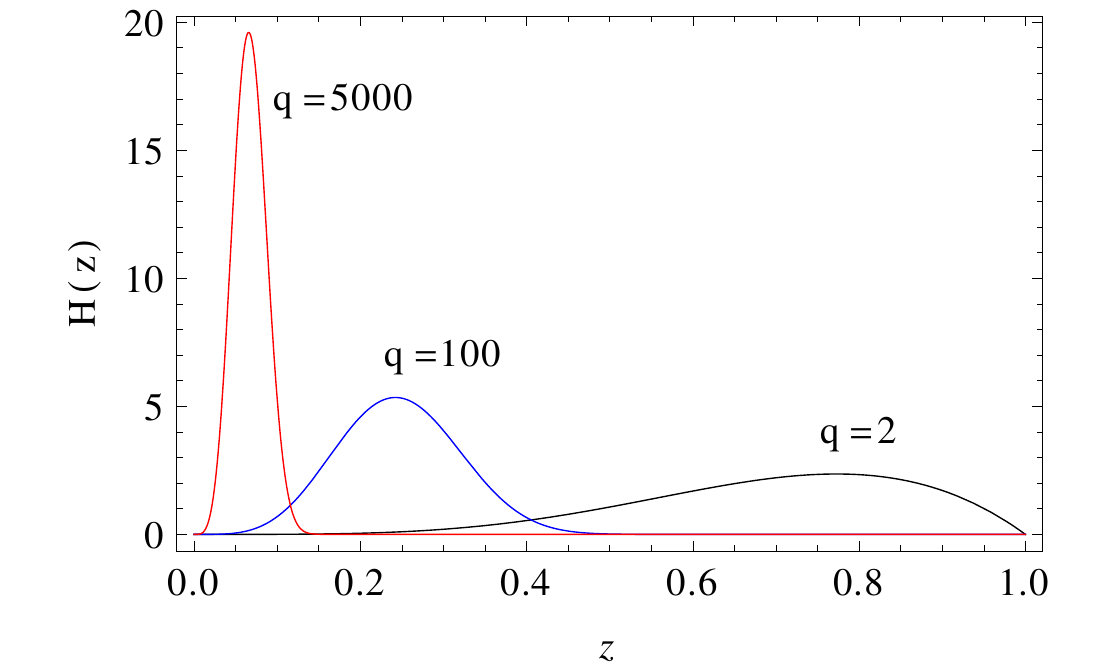}
\caption{The linearized tensor perturbation for three different choices of $q$. As $q$ increases the function becomes squeezed and pushed towards $z=0$. These functions correspond to the lowest overtone number, and have been normalized so that the area under the curve is 1.}
\end{figure}
\begin{figure}[H]
\centering
\includegraphics[scale=.8]{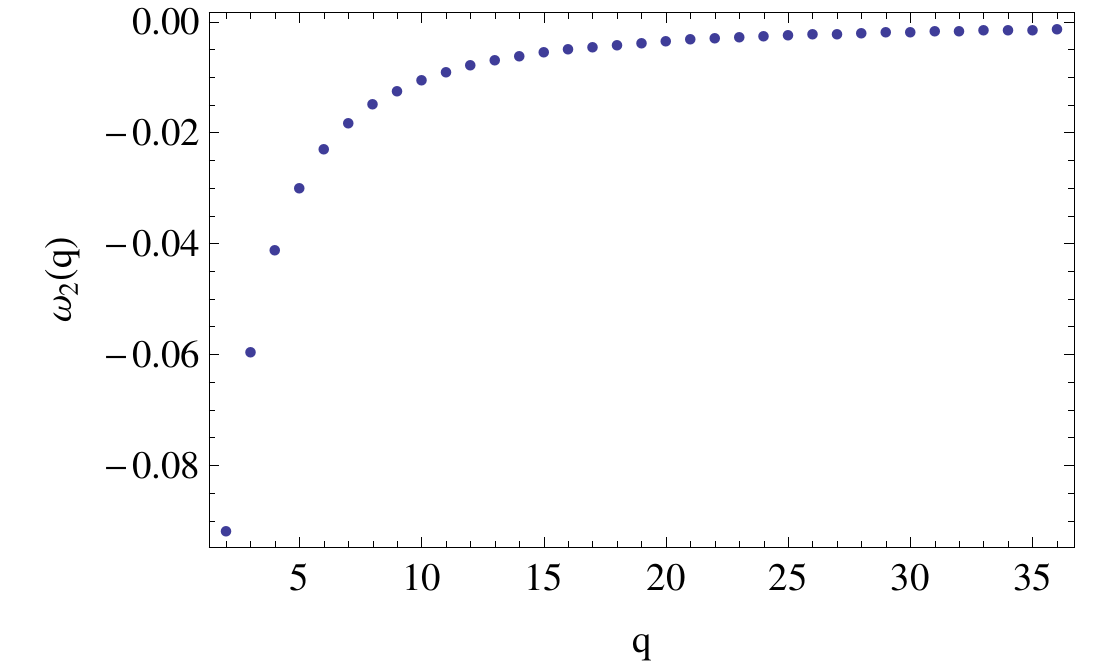}
\caption{Plot of the second order change in  frequency, $\omega_2(q)$. For all $q$ studied, $\omega_2 < 0$. Normalization is fixed so that  $P = \epsilon^2 V_2$ is constant.}
\end{figure}

\section{Geons and Holographic Superconductors}
\subsection{AdS Soliton Superconductors}
In this section we briefly review the holographic superconductor in the AdS soliton background \cite{Nishioka, HorowitzWay}. This system has many features in common with the dimensionally reduced geons, and the review will be useful when we investigate the possibility that geons could model holographic superconductors. We also present a simple characterization of the superconducting phase transition and show that it is {\it not} caused by a linear instability towards scalar condensation.

Consider the Einstein-Maxwell-scalar theory defined by the following action
\be 
S = \int d^5x \sqrt{-g}\Big(R+\frac{12}{L^2} -\frac{1}{4}F^2-|(\nabla -i q A) \Psi|^2- m^2 |\Psi|^2 \Big).
\ee
The possibility that this theory could describe a $T=0$ insulator/superconductor phase transition in the background geometry of the AdS soliton was first considered in Ref.\ \cite{Nishioka}. They worked in the probe limit in which the backreaction of the metric is neglected but the nonlinearities of the Maxwell and scalar fields is preserved. A superconducting phase transition was found to occur as the chemical potential $\mu$ was increased from zero to some critical value, $\mu_c$. Then \cite{HorowitzWay} considered the full backreaction and found the complete phase diagram. Both groups  either explicitly or implicitly considered the following ansatz for the matter fields
\be 
A=\phi(r)dt, \qquad \Psi = \psi(r)e^{-i\omega t}.
\ee
As a result of gauge invariance, the equations of motion are invariant under the transformation
$ \phi \rightarrow \phi + C,$ $\omega \rightarrow \omega - q C, $
where $C$ is a constant. 
In both papers $C$ was chosen to set $\omega=0$. Since the chemical potential is defined to be the leading term in the near-boundary expansion of the gauge field, $\phi = \mu - \rho/2r^2 + \mathcal{O}(r^{-4})$, this corresponds to a choice of chemical potential. 

The critical chemical potential found in the probe limit in Ref.\ \cite{Nishioka} simply corresponds to the smallest normal mode frequency of a charged scalar field in the AdS soliton background (divided by the charge $q$). To see this, consider the system in the probe limit right at the critical point, $\phi = \mu_c$. The Maxwell field is pure gauge, and the scalar field is negligibly small, so the equations of motion reduce to a single linear ODE for the static scalar field $\Psi = \psi(r)$. By using the above symmetry this field configuration can be transformed to $\phi = 0$, $\Psi = \psi(r)\exp(-iq\mu_c t)$. The $\phi$ equation of motion is trivially satisfied, and the $\psi$ equation is identical to that determining the normal modes, except with $\omega\rightarrow q\mu_c$. So the determination of $\mu_c$ is equivalent to the problem of finding the normal mode frequencies of a scalar field in the AdS soliton background. This connection between the critical chemical potentials and the normal mode frequencies was first discussed in $\cite{Cai}$.

There is a crucial difference between the superconducting phase in the AdS soliton background and the original holographic superconductor based on the Schwarzschild AdS solution \cite{HHH}. In the latter case, below a critical temperature, the black hole becomes unstable to forming charged scalar hair. In the AdS soliton there is no instability for any value of the chemical potential.  Instead, the insulator/superconductor phase transition reflects  which configuration has lower free energy and hence dominates a grand canonical ensemble. If there were an unstable linearized mode growing exponentially in time for some $\phi = \mu > \mu_c$, then the above shift symmetry could be used to set $\phi = 0$ at the cost of shifting the real part of $\omega$. This shift would not affect the exponentially growing behaviour of the mode, so it  would therefore be an unstable perturbation of the AdS soliton. But it is known that the AdS soliton is stable to linearized perturbations of both the metric and a free massless scalar field \cite{HorowitzMyers,ConstableMyers}. So such a growing linearized mode cannot exist, and the appearance of a new branch of static solutions (the superconducting condensates) is not related to any linear instability.

\subsection{Holographic Dual of the Geons}
We now investigate the possibility that the dimensional reduction of the geons could model a holographic superconductor.
Recall the standard Kaluza-Klein reduction of a five dimensional metric with a $U(1)$ symmetry. One can write the metric in the form
\be 
ds^2 = g_{AB}dx^Adx^B = \phi^2 (dy + A_a dx^a)^2 + \frac{1}{\phi} g_{ab}^{(4)} dx^a dx^b,
\ee
where $x^a = (r,x^{\mu})$ and the functions depend on the $x^a$ coordinates only. This parametrization leads to a four-dimensional Einstein-Maxwell-dilaton theory in the Einstein frame. The off-diagonal part of the metric, $A_a$, is interpreted as a Maxwell field. Five-dimensional coordinate transformations of the form $y \rightarrow y + \lambda(x)$ correspond to four-dimensional gauge transformations, $A_a(x) \rightarrow A_a(x) + \partial_a \lambda(x)$. 

The AdS soliton can, of course, be written in the above form\footnote{The four dimensional Einstein metric will be singular at the location of the tip and is not asymptotically AdS. However in terms of applications to gauge/gravity duality, both of these problems can be resolved by applying holography directly to the five dimensional solution \cite{Skenderis}.}  with $A_a = 0$. A simple five-dimensional coordinate transformation $y\rightarrow y + \mu t$ generates a nonzero gauge field $A_t = \mu$.   If there is a phase transition to forming the geon at a critical value $\mu = \mu_c$, then near $\mu_c$ the amplitude of the geon will be small and it can be approximated by the leading ${\cal O}(\epsilon)$ correction to the metric (\ref{graviton}). When $\mu = \omega/q$ (where $\omega$ is the frequency of the linearized graviton mode) the dimensional reduction  of this mode corresponds to a static, charged spin-2 field. In the dual boundary theory, this corresponds to turning on a charged spin-2 condensate $\langle {O}_{x_1x_2} \rangle$. Since the AdS soliton is stable to linearized metric perturbations, the appearance of this new branch of static solutions is analogous to turning on the superconducting condensate in Ref.'s \cite{Nishioka,HorowitzWay}; in neither case is the condensation the result of a linearized instability.

The key question is whether the condensate has lower free energy than the original state with no condensate. To compute this, one needs to construct the geon to higher order as we have done in the previous section. Consider the five-dimensional theory in the zero temperature grand canonical ensemble. The relevant free energy functional is the Gibbs free energy, which, in the absence of any black hole horizons or true five-dimensional Maxwell fields, takes the form $G = E-(\omega/q) P$. Making use of the first law (\ref{firstlaw}), 
 the Gibbs free energy obeys 
\be
dG = -\frac{P}{q}d\omega = -\frac{2P_2\omega_2}{q} (1 + \mathcal{O} (\epsilon^2)) V_2 \epsilon^3 d\epsilon.
\ee
Since $P_2 > 0$, the sign of the free energy depends on the sign of $\omega_2$, which we have found is always negative (see Fig. 2). Therefore, at least in perturbation theory, the geon will always have a larger free energy than the AdS soliton. Since the free energy is unaffected by dimensional reduction, we conclude that the spin-2 condensate will also have a larger free energy than the field theory state with no condensate. This result spoils the hope that the perturbative geon could model a $d$-wave superconductor.

What is the key difference between this case and the earlier result that there is a phase transition when the Maxwell field and charged scalar field are added to the five dimensional action? The most likely explanation is that when a charged scalar is added, one can increase the charge $q$ keeping the mass fixed, while in the Kaluza-Klein case, increasing $q$ also increases the effective mass in the four dimensional theory. Indeed, it was shown in \cite{HorowitzWay} that for $m^2 = -15/4$ and $q=1$ the free energy increases when the scalar turns on. It decreases only for larger $q$ (with the same $m$). However, even for $q=1$, it turns out that as the amplitude of the scalar field increases, the change in free energy eventually becomes negative and there is a first order phase transition \cite{HorowitzWay}.

It remains possible that the exact geon solutions will behave like the $q=1$ charged scalar. However, even if the change in free energy eventually becomes negative, the condensate will not be pure spin-2. From the structure of the perturbative Einstein equations, it is clear that the first order seed perturbation $H_1^{(0)}(z)$  sources an infinite number of higher Fourier modes, and that the higher order perturbations are complicated combinations of scalar, vector, and tensor modes. Therefore, the corresponding state in the dual field theory is a nonlinear combination of spin-0, 1, and 2 condensates of various charges. This  is conveniently described in terms of $\langle T_{\mu\nu}\rangle$.  Since the metric is the only nonzero bulk field in five dimensions, the only dual operator with a nonzero expectation value is the (traceless) stress tensor. The Maxwell field on the boundary arises from Kaluza-Klein reduction of the (fixed) boundary metric.

To summarize, we have perturbatively constructed a class of geons in the background geometry of the AdS soliton to third order. We only considered geons seeded by tensor perturbations; vector and scalar perturbations would lead to different classes of geons. These solutions have an exact helical Killing vector $K = \partial_t + (\omega/q) \partial_y$. We considered the dimensional reduction of the geons and found some features suggestive of $d$-wave superconductors. However, these geons do not provide a gravitational dual of a continuous phase transition to a  superconductor.
 
We have also seen that the previously studied phase transition in the AdS soliton is not the result of a linear instability. The dimensionally reduced spin-2 condensate we construct is also not the result of a linear instability. Perhaps the Kaluza-Klein approach towards holographic superconductivity would be more successful if the gravitational background became linearly unstable to a metric perturbation as some control parameter is varied. This possibility deserves further investigation.

\vskip 1cm
\centerline{\bf Acknowledgements}
\vskip 1cm
We would like thank J. Santos and B. Way for useful discussions. We are especially grateful to J. Santos for suggesting this project. This work was supported in part by NSF grant PHY12-05500.




\vskip 3cm
\begin{thebibliography}{99}

\bibitem{BrillHartle}
  D.~R.~Brill and J.~B.~Hartle,
  ``Method of the Self-Consistent Field in General Relativity and its Application to the Gravitational Geon,''
  Phys.\ Rev.\  {\bf 135} (1964) B271.

\bibitem{Louko}
  J.~Louko, R.~B.~Mann and D.~Marolf,
  ``Geons with spin and charge,''
  Class.\ Quant.\ Grav.\  {\bf 22} (2005) 1451
  [arXiv:0412012 [gr-qc]].

\bibitem{Dias1}
  O.~J.~C.~Dias, G.~T.~Horowitz and J.~E.~Santos,
  ``Gravitational Turbulent Instability of Anti-de Sitter Space,''
  Class.\ Quant.\ Grav.\  {\bf 29} (2012) 194002
  [arXiv:1109.1825 [hep-th]].
  
\bibitem{Bizon}
  P.~Bizon and A.~Rostworowski,
  ``On weakly turbulent instability of anti-de Sitter space,''
  Phys.\ Rev.\ Lett.\  {\bf 107} (2011) 031102
  [arXiv:1104.3702 [gr-qc]].

\bibitem{Witten}
  E.~Witten,
  ``Anti-de Sitter space, thermal phase transition, and confinement in gauge theories,''
  Adv.\ Theor.\ Math.\ Phys.\  {\bf 2} (1998) 505
  [arXiv:9803131 [hep-th]].

\bibitem{HorowitzMyers}
  G.~T.~Horowitz and R.~C.~Myers,
  ``The AdS/CFT correspondence and a new positive energy conjecture for general relativity,''
  Phys.\ Rev.\ D {\bf 59} (1998) 026005
  [arXiv:9808079 [hep-th]].
    
\bibitem{Nishioka}
  T.~Nishioka, S.~Ryu and T.~Takayanagi,
  ``Holographic Superconductor/Insulator Transition at Zero Temperature,''
  JHEP {\bf 1003} (2010) 131
  [arXiv:0911.0962 [hep-th]].
  
\bibitem{HorowitzWay}
  G.~T.~Horowitz and B.~Way,
  ``Complete Phase Diagrams for a Holographic Superconductor/Insulator System,''
  JHEP {\bf 1011} (2010) 011
  [arXiv:1007.3714 [hep-th]].
  
\bibitem{HHH}
  S.~A.~Hartnoll, C.~P.~Herzog and G.~T.~Horowitz,
  ``Building a Holographic Superconductor,''
  Phys.\ Rev.\ Lett.\  {\bf 101} (2008) 031601
  [arXiv:0803.3295 [hep-th]].

\bibitem{GubserPufu}
  S.~S.~Gubser and S.~S.~Pufu,
  ``The Gravity dual of a $p$-wave superconductor,''
  JHEP {\bf 0811} (2008) 033
  [arXiv:0805.2960 [hep-th]].
  
\bibitem{Chen}
  J.~-W.~Chen, Y.~-J.~Kao, D.~Maity, W.~-Y.~Wen and C.~-P.~Yeh,
  ``Towards A Holographic Model of $D$-Wave Superconductors,''
  Phys.\ Rev.\ D {\bf 81} (2010) 106008
  [arXiv:1003.2991 [hep-th]].
  
\bibitem{Benini}
  F.~Benini, C.~P.~Herzog, R.~Rahman and A.~Yarom,
  ``Gauge gravity duality for $d$-wave superconductors: prospects and challenges,''
  JHEP {\bf 1011} (2010) 137
  [arXiv:1007.1981 [hep-th]].
  
\bibitem{Kim} 
K.~Y.~Kim, K.~Skenderis and M.~Taylor,
in preparation.
  
\bibitem{ConstableMyers}
  N.~R.~Constable and R.~C.~Myers,
  ``Spin two glueballs, positive energy theorems and the AdS/CFT correspondence,''
  JHEP {\bf 9910} (1999) 037
  [arXiv:9908175 [hep-th]].
 
\bibitem{Gubser}
  S.~S.~Gubser,
  ``On nonuniform black branes,''
  Class.\ Quant.\ Grav.\  {\bf 19} (2002) 4825
  [arXiv:0110193 [hep-th]].

\bibitem{Balasubramanian}
  V.~Balasubramanian and P.~Kraus,
  ``A Stress tensor for Anti-de Sitter gravity,''
  Commun.\ Math.\ Phys.\  {\bf 208} (1999) 413
  [arXiv:9902121 [hep-th]].
  
\bibitem{Wald:1993ki}
  R.~M.~Wald,
  ``The First law of black hole mechanics,''
  in {\it Directions in general relativity, vol. 1} (B. Hu and T. Jacobson, eds.), Cambridge University Press (1993) 358
  [arXiv:9305022 [gr-qc]].
  
\bibitem{Cai}
  R.~-G.~Cai, X.~He, H.~-F.~Li and H.~-Q.~Zhang,
  ``Phase transitions in AdS soliton spacetime through marginally stable modes,''
  Phys.\ Rev.\ D {\bf 84} (2011) 046001
  [arXiv:1105.5000 [hep-th]].

\bibitem{Skenderis}
  B.~Gouteraux, J.~Smolic, M.~Smolic, K.~Skenderis and M.~Taylor,
  ``Holography for Einstein-Maxwell-dilaton theories from generalized dimensional reduction,''
  JHEP {\bf 1201} (2012) 089
  [arXiv:1110.2320 [hep-th]].

  
  
\end{thebibliography}
\end{document}